\def\t{\textstyle}

\vglue 3cm
\baselineskip=20pt
\centerline{\bf CLASSIFICATION OF INTEGRABLE EVOLUTION EQUATIONS  }
\centerline{\bf OF THE FORM $u_t=u_{xxx}+f(t,x,u,u_x,u_{xx})$}

\vskip 2cm
\baselineskip=14pt
\centerline {Ay\c se H\"umeyra Bilge$^{(*)}$}
\vskip .3cm
\centerline {Department of Mathematics, Faculty of Sciences,}
\centerline {Bilkent University}
\centerline {06533 Ankara, TURKEY}

\vfill

\baselineskip=16pt
\centerline{\bf Abstract.}

We obtain the classification of integrable equations of the form
$u_t=u_{xxx}+f(t,x,u,u_x,u_{xx})$ using the formal symmetry method of Mikhailov
et al [A.V.Mikhailov, A.B.Shabat and V.V.Sokolov, in {\it What
is Integrability} edited by V.E. Zakharov (Springer-Verlag, Berlin
1991)]. We show that all such equations can be transformed to an
integrable equation of the form $v_t=v_{xxx}+f(v,v_x,v_{xx})$ using
transformations $\Phi(x,t,u,v,u_x,v_x)=0$, and the
$u_{xx}$ dependence can be eliminated except for two equations.

\vskip 2cm
\vskip .2cm
\leftline {Permanent Address:}
\leftline {Department of Mathematics}
\leftline {TUBITAK Marmara Research Center}
\leftline {P.O.Box 21, 41470 Gebze, Kocaeli TURKEY}
\leftline {e-mail: bilge@yunus.mam.tubitak.gov.tr}

\eject

\baselineskip=20pt
\vglue 2cm
\noindent
{\bf 1. Introduction.}

The existence of ``formal symmetries" has been proposed as an integrability
test by Mikhailov et al [1].
 A  formal symmetry is a truncated expansion of the recursion operator in
inverse powers of
$D={\partial \ \over \partial x}$, hence the equations admitting
a formal symmetry are candidates for those equations admitting a recursion
operator.
The classifications of evolution equations
$u_t=F(x,u,u_x,u_{xx},u_{xxx})$  (except for a class essentially nonlinear in
$u_{xxx}$), and
$u_t=h(x,t)u_3+f(x,t,u,u_x)$ are given respectively in [1] and
[2]. In this paper
we obtain  the classification of
evolution equations
$u_t=u_{xxx}+f(t,x,u,u_x,u_{xx})$
using the existence of canonical conserved densities $\rho_i$,
$i=0,\dots ,3$.
We show that all such equations
can be transformed to an equation without explicit $x$ and $t$
dependence using first order differential substitutions.  The list of
integrable equations given in [1] is further
simplified by the availibility of time dependent transformations.
Furthermore, the  $u_{xx}$ dependence is eliminated, but two of the transformed
equations are nonlocal. Thus, up to first order differential
substitutions, there are only three equations in the class considered.

\vskip .5cm
\noindent
{\bf 2. Formal symmetries, recursion operators and differential substitutions.}
\vskip .3cm

Let $u_t=F[u]$ be an evolution equation where $F$ is a differential
polynomial. A {\it symmetry}  $\sigma$ of this equation is a
differential polynomial which satisfies
the linearized equation $\sigma_t=F'\sigma$ where $F'$ is the Frechet
derivative of $F$. The {\it recursion operator} $R$ is a linear operator that
sends symmetries to symmetries, i.e. $R\sigma$ is a symmetry whenever $\sigma
$ is a symmetry [3]. In general the recursion operator
$R$ is an integro-differential operator, and satisfies
$(R_t+[R,F'])\sigma=0$. If the symmetries are dense in the solutions of this
equation, then $R$ can be defined as the solution of the operator equation
$$R_t+[R,F']=0.\eqno(2.1)$$
In $1+1$ dimensions the integral terms in the  recursion operator can be
expanded in inverse powers of $D={d\over dx}$, and Eq.(2.1)
can be solved recursively, leading to linear first
order differential equations for the coefficients of $R$. The solvability of
these equations in the class of local functions gives certain conserved density
conditions, $(\rho_i)_t=D\eta_i$. These conserved density conditions in turn
give
nonlinear  partial differential equations for $F$, whose solutions lead to the
classification.

We consider the classification of evolution equations
$$u_t=u_3+f(x,t,u,u_1,u_2).\eqno(2.2)$$
In the following,
subscripts denote differentiation with respect to $x$.
The classification is obtained  using the existence of
conserved densities $\rho_0,\dots , \rho_3$, and we have chosen the coordinate
transformations so as to put the equations in the form given in [1].

We shall use the following general result for the elimination of $u_2$.
Consider the evolution equation
$u_t=u_m+f(u_{m-1}, \dots ,u,x,t)$.
If this
equation is integrable then
$\rho_0=\partial f/ \partial u_{m-1}$ is a conserved density, hence $\int
\rho_t$ is a local function. The dynamical variable $s$ defined by
$s=u_{m-1}e^{{1\over m}\int\rho_0}$,
satisfies an equation of the form
$s_t=s_m+g(s_{m-2}\dots ,s,u,\dots u_{m-1})$
i.e. $s_{m-1}$ is eliminated.
If $\rho_0$ is itself a total derivative, then $s$ is a local function.
Otherwise, one has to introduce a new dynamical variable $v=\int \rho_0$, and
apply the transformation to the new equation.
This transformation of course
introduces non-locality, and the
the crucial point is to ensure that the transformed equation is local.
We shall  also use the coordinate transformation
$x\to x+h(t)t$ to eliminate the term $h'(t)u_1$.
Other differential substitutions will be described as needed.

\vskip .3cm
\noindent
{\bf 3. Classification results.}
\vskip .2cm

For equations of the form (2.2), the first
four nontrivial conserved density conditions are given below. We shall obtain
a classification based on the existence of these conserved densities.

$$\eqalignno{
\rho_0=&{\partial f\over u_2},&(3.1a)\cr
\rho_1=&{1\over 3}\left({\partial f\over u_2}\right)^2-
{\partial f\over \partial u_1},&(3.1b)\cr
\rho_2=&{\partial f\over \partial u}+{2\over 27}
\left({\partial f\over u_2}\right)^3-{1\over 3}
{\partial f\over \partial u_2}
{\partial f\over \partial u_1}+{1\over 3}\eta_0,&(3.1c)\cr
\rho_3=&\eta_1,&(3.1d)\cr
}$$
where
$(\rho_i)_t=D\eta_i$, $i=0,1$.

The method is to compute the time derivative of the $\rho_i$'s and integrate by
parts until a term that involves the highest derivative non-linearly appears.
Then
the coefficient of this highest derivative has to be set to zero, and we obtain
partial differential equations  for  $f$.

As a first step, it can easily be seen that  $f$ has to be  a third order
polynomial in $u_2$, hence, we obtain the $u_2$ dependence
of $f$, and the evolution equation has the form
$$u_t=u_3+A(x,t,u,u_1)u_2^2+B(x,t,u,u_1)u_2+C(x,t,u,u_1).\eqno(3.2)$$
Then, we
 obtain the following nonlinear differential equation for the $u_1$
dependence of $A$,
$${\partial^2 A\over  \partial u_1^2}-4
  {\partial   A\over  \partial u_1  }A+{16\over 9} A^3=0.\eqno(3.3)$$
For  $A\ne 0$,   the solution is given by
  $$A=-{3\over 4}{\partial Z/ \partial u_1\over Z},\quad \quad
{\partial ^3Z\over \partial u_1^3}=0.\eqno(3.4)$$
Hence
$$Z=a(x,t,u)u_1^2+b(x,t,u)u_1+c(x,t,u).\eqno(3.5)$$
In Sections 4,  5 and 6 we shall study respectively the cases where $A=B=0$,
$A=0$, $B\ne 0$ and $A\ne 0$.

\vskip .5cm
\noindent
{\bf 4. Classification for the case  $A=B=0$.}
\vskip .2cm

We will show that the equations in this class consist of the KdV, mKdV, their
potential forms and the Callegero-Degasperis-Fokas equation.
Actually, the
classification of integrable equations of the form
$u_t=h(x,t)u_{xxx}+f(t,x,u,u_x)$
has been obtained in [2], hence
we will only give an outline of the derivations for
completeness. The use of the coordinate transformations (4.2) given below is
crucial in obtaining the classification.
Using the conserved density conditions we first obtain
$$u_t=u_3-{1\over 2}k(t)^2u_1^3
+a_2(x,t,u)u_1^2+a_1(x,t,u)u_1+a_0(x,t,u).\eqno(4.1)$$
The first branching is given by
$k(\partial a_2/ \partial u)=0$.
The coordinate transformations leaving the form of (4.1) invariant are
$\tilde{u}=\alpha(t)u+\beta(x,t)$, and
$$\eqalignno{ \tilde{u}_t=&\tilde{u}_3-{1\over 2}{k^2\over
\alpha^2}\tilde{u}_1^3+
\tilde{u}_1^2
 \left[{3\over 2}{\beta _x\over \alpha^2}k^2+{1\over \alpha}a_2\right]
+\tilde{u}_1\left[
-{3\over 2} {\beta_x^2\over \alpha^2}k^2-2{\beta_x\over
\alpha}a_2+a_1\right]\cr
&
-{\alpha_t\over \alpha}(\tilde{u}-\beta)+\beta_t-\beta_{xxx}
+{1\over 2} {\beta_x^3\over \alpha_2}k^2+{\beta_x^2\over \alpha}a_2
-\beta_x a_1+a_0\alpha. &(4.2)\cr}$$
For $k\ne0$, we set $k=1$, $a_2=0$, and using the conserved density conditions
we  obtain the
Callegero-Degasperis-Fokas
equation
 $$u_t=u_3-{1\over 2}u_1^3+u_1[c_1e^{2u}+c_2e^{-2u}].\eqno(4.3)$$
 For $k=0$, we first obtain
  $u_t=u_3+a_2(x,t,u)u_1^2+a_1(x,t,u)u_1+a_0(x,t,u),$
where $a_2=b_0(x,t)$,
$a_1=c_2(x,t)u^2+c_1(x,t)u+c_0(x,t)$,
 $a_0=d_3(x,t)u^3+d_2(x,t)u^2+d_1(x,t)u+d_0(x,t)$. The branching is
given by the condition $b_0c_2=0$.
For $b_0\ne 0$ we have the potential KdV equation,
$$u_t=u_3+u_1^2,\eqno(4.4) $$
For $b_0=0$, using the conserved density conditions and appropriate
coordinate transformations we obtain the KdV
$$u_t=u_3+uu_1\eqno(4.5)$$
and mKdV equations
$$u_t=u_3+u^2u_1.\eqno(4.6) $$
It is well known that all the equations in this class are related via Miura
transformations, hence up to first order differential
substitutions there is a single equation in the class $u_t=u_3+f(x,t,u,u_x)$.

      \vskip .5cm
\noindent
{\bf 5. Classification for the case $A=0 $, $B\ne 0$.}
      \vskip .2cm

      We will show that all equations in this class are linearizable.
The conserved density conditions imply that the evolution
equation  has to be in the form
$$u_t=u_3+b_1(x,t,u)u_1u_2+b_0(x,t,u) u_2
+c_3(x,t,u) u_1^3 +c_2(x,t,u) u_1^2 +c_1(x,t,u) u_1+c_0(x,t,u).\eqno(5.1)$$
Then the transformation $u\to \phi(u) $ gives
$b_1\to {b_1\over \phi'}-3{\phi''\over \phi'^2}$ and we can set $b_1=0$. Then
the allowable transformations are of the form $u\to
\alpha(x,t)u+\beta(x,t)$. The conserved
density conditions give $\partial c_3/\partial u=0$, and we set $c_3$ constant
by choosing $\alpha$. Then we have the branching $b_0c_3=0$. If $b_0=0$, (5.1)
is independent of $u_2$, hence we set $c_3=0$. Then we have
$$\eqalignno{
b_0=&b_{02}(x,t)u^2+b_{01}(x,t)u+b_{00}(x,t),\cr
c_2=&c_{21}(x,t)u+c_{20}(x,t),\cr
c_1=&{1\over 3}b_{02}^2u^4+{2\over 3}b_{01}b_{02}u^3
  +c_{12}(x,t)u^2+c_{11}(x,t)u+c_{10}(x,t).&(5.2)\cr}$$
We have two subcases depending on whether $b_{02}=0$ or $b_{02}\ne 0$.

\vskip .3cm
\noindent
{\bf The case $b_{02}\ne 0$:} In this case we use coordinate
transformations
$u\to \alpha(x,t)u+\beta(x,t)$ to set $b_{02}=3$ and $b_{01}=0$. Then we obtain
the following equation.
$$
u_t= u_3+3u^2u_2+9uu_1^2+3u^4u_1+c_{10}(x,t)u_1+{1\over
2}(c_{10})_xu.\eqno(5.3)$$
This equation can be linearized by the following sequence of Miura and
coordinate
transformations. Recall that the transformation (2.5) eliminates $u_2$ but
in this case $\int \rho_0$ is not a local function. Thus we first use the
potentiation  $u_1\to u^2$ to obtain
$$u_t=u_3-{3\over 4} {u_2^2\over u_1}+3 u_2
u_1+u_1^3+u_1c_{10}(x,t)+d(t).$$
Then using the transformation $u\to e^{2 u}$ we obtain
$$u_t=u_3-{3\over 4} {u_2^2\over u_1}+c_{10}(x,t)u_1+2 d(t)u.$$
Finally the transformation $u\to  2\sqrt{u_1} $ we  obtain the linear equation
$$u_t=u_3+c_{10}(x,t)u_1+{1\over 2}u((c_{10})_x+2d).\eqno(5.4)$$

\vskip .3cm
\noindent
{\bf The case $b_{02}= 0$:} In this case we use point transformations
to set $b_{01}=3$ and $b_{00}=0$. Then the conserved density conditions give
$$u_t=u_3+3uu_2+3 u_1^2+3 u^2u_1 +u_1c_{10}(x,t)+(c_{10})_xu+c_{00}(x,t).
\eqno(5.5)$$
The potentiation $u_1\to u$ gives
$$u_t=u_3+3u_1u_2+u_1^3+u_1 c_{10}(x,t)+\int c_{00}(x,t)\ dx.$$
Then the point transformation
$u\to e^u$ gives the linear equation
$$u_t=u_3+u_1 c_{10}+u\int c_{00} dx.\eqno(5.6)$$

For completeness, we give the interacting soliton equation of
the  KdV equation that we would obtain if we  had taken $c_2=0$ instead of
$b_1=0$ in (5.1)
$$
u_t=u_3-3{u_2u_1\over u}+{3\over 2} {u_1^3\over u^2}+{3\over 2}m
u_1.\eqno(5.7)$$

\vskip .3cm
\noindent
{\bf 6. Classification for the case $A\ne 0$.}

When $A\ne 0$, the coordinate transformation $u\to \phi(x,t,u)$ results in
$$A\to (2a u_1+\phi_u b-2\phi_x a)
\big[au_1^2 +u_1(\phi_u b-2\phi_x a)+(\phi_u^2c-\phi_u\phi_xb+\phi_x^2a)
\big]^{-1}.\eqno(6.1)$$
Thus for $a=0$, these coordinate transformations can be used to take $b=1$,
$c=0$, similarly for $a\ne 0$,  we can take $a=1$ and $b=0$. The cases $c=0$
and $c\ne 0$ has to be considered separetely in the last subclass. As a result
$A$ has the forms studied in Sections 6.1-3 below.
In all cases the allowable cordinate transformations should satisfy
$\phi_x=0$.
We will use the transformation
$$v=\int e^{{1\over 3}\int\rho_0}u_2\eqno(6.2)$$
to eliminate $u_2$ term.

\vskip .2cm
\noindent
{\bf 6.1 Classification for the case $A=-{3\over 4}u_1^{-1}$.}

We will show that all equations in this class are either linearizable, or
transformable to the mKdV equation. We first obtain
$$B=a_{11}(x,t,u)u_1^{1/2}+a_{12}(x,t,u)u_1+a_{13}(x,t,u).$$
Then $C$ can be obtained by solving a fourth order o.d.e., and
if $a_{13}$ is nonzero, it has $u_1^2ln(u1)$ and $u_1ln(u_1)$ terms.
The first branching is given by the condition
$a_{11}a_{13}=0$. If we assume $a_{11}=0$, $a_{13}\ne 0$, the conserved
density conditions give $a_{13}=0$. Hence we assume
$a_{11}\ne 0$.

\noindent
{\bf The case $a_{11}\ne 0 $ (linearizable equation):}
We will  show that the equation with $a_{11}\ne 0$ is linearizable.
We have
$$\eqalignno{
C=&a_{01}(x,t,u)u_1^{3/2}
+a_{02}(x,t,u)u_1^2+a_{03}(x,t,u)u_1+a_{04}(x,t,u)\cr
&\quad+ \left[{2\over 3}(a_{12})_u+{1\over 9}a_{12}^2\right]u_1^3
+{2\over3}a_{11}a_{12}u_1^{5/2}.\cr}$$
The conserved density conditions give $(a_{04})_x=0$, and we set $a_{04}=0$ by
a coordinate transformation. As a result for $a_{11}\ne 0$, we obtain  the
following equations.
 $$\eqalignno{
(a_{11})_t=&(a_{11})_{xxx}+(a_{11})_xa_{03}+{1\over 2} (a_{03})_xa_{11},\cr
(a_{12})_t=&-{2\over 3}(a_{11})_{xx}a_{11}+{1\over 3}(a_{11})_x^2
              -{1\over 3} a_{11}^2a_{03},\cr
(a_{11})_u=&{1\over 3} a_{11}a_{12},\quad\quad
(a_{12})_x=-{1\over 3} a_{11}^2,\quad\quad
(a_{03})_u={2\over 3} (a_{11})_xa_{11},\cr
a_{01}=&2 (a_{11})_x,\quad\quad
a_{02}= 0,\quad\quad
a_{04}= 0.&(6.3)\cr}$$
We make the differential substitution
$s=\psi(x,t,u,u_1)$
which gives an equation where $s_2^2$ term is not present. Elimination of  the
coefficient of $s_2s_1$ requires that
$$\psi=\psi_1(x,t,u)\sqrt{u1}+\psi_0(x,t,u).$$
Then  $s_2s^2$ and $s_2$ are eliminated by choosing
$\psi_1$ and $\psi_0$ as solutions of
$$3(\psi_1)_u-a_{12}\psi_1=0,\quad\quad
  3(\psi_1)_x+a_{11}\psi_0=0.\eqno(6.4)$$
The resulting equation for $s$ is
independent of $u$ by virtue of the equations (6.3) and (6.4) and it can be
transformed
to equation in the previous section by a linear coordinate transformation, and
then it is linearized.

\noindent
{\bf The case $a_{11}=0$ (equations linearizable or transformable to mKdV):}
When $a_{11}=0$,
 we have $(a_{12})_x=0$, and we can make
$a_{12}=0$ by a point transformation, hemce $B=0$. Then we are allowed to make
transformations of the form $u\to \alpha(t)u+\beta(t)$. The conserved density
conditions give two equations in this class, the first one being a constant
coefficient equation.
$$\eqalignno{
u_t&= u_3-\t {3\over 4} u_1^{-1}u_2^2+a_{01}u_1^{3/2}+a_{02}u_1^2+a_{03}
u_1+a_{04},&(6.5a)\cr
u_t&= u_3-\t {3\over 4} u_1^{-1}u_2^2+a_{03}(x,t) u_1.&(6.5b)\cr}$$

We will show that both equations are transformable to an equation independent
of $u_2$. The conserved density
$\rho_0=-{3\over 4}{u_2\over u_1}$ is trivial, the transformation (6.2)
reduces to $v=2\sqrt{u_1}$, and we have
$$u_t-u_3+{3\over 4}{u_2^2\over u_1}\to v_t-v_3.          $$
Thus we  obtain
respectively the
following equations
$$\eqalignno{
v_t=&v_3+v_1[{1\over 2}a_{02}v^2+{3\over 4}a_{01}v +a_{03} ],&(6.6a)\cr
v_t=&v_3+a_{03}v_1+{1\over 2}(a_{03})_x v.&(6.6b)\cr}$$
The second equation is linear and the first one is transformable to the
mKdV equation.
Therefore, all integrable  equations with $A=-{3\over 4}u_1^{-1}$ are either
linearizable, or transformable to the mKdV equation.

\vskip .2cm
\noindent
{\bf 6.2 Classification for the case $A=-{3\over 2}u_1^{-1}$.}

In this case, using the conserved density conditions we obtain
$$u_t=u_3-{3\over 2}{u_2^2\over u_1}+a_{01}(t,u){1\over u_1}
+a_{02}(t,u)u_1^3+a_{03}u_1+a_{04}(t,u),\eqno(6.7)$$
where the coefficients have to satisfy the following equations.
$$\eqalignno{
(a_{01})_t=&-(a_{01})_ua_{04}+2(a_{04})_ua_{01},&(6.8a)\cr
(a_{02})_t=&-(a_{02})_ua_{04}-(a_{04})_{uuu}-2(a_{04})_ua_{02},&(6.8b)\cr
         0=&(a_{01})_{5u}+10(a_{01})_{uuu}a_{02}+15(a_{01})_{uu} (a_{02})_u\cr
           &\ \ +a_{01})_u[9(a_{02})_{uu}+16a_{02}^2]
+2 a_{01}[(a_{02})_{uuu}+8(a_{02})_ua_{02}].&(6.8c)}$$
The equations (6.8a-c) form a consistent system, hence (6.7) is an integrable
equation.
The conserved density $\rho_0=-{3\over 2}{u_2\over u_1}$ is trivial and
the transformation $v={\rm ln}\ u_1$  gives
$$u_t-u_3+{3\over 2}{u_2^2\over u_1}\to v_t-v_3+{1\over 3} v_1^3.\eqno(6.9)$$
We will show that, generically (6.7) cannot be transformed to a {\it local}
equation independent of $u_2$.
  The coordinate transformation $u\to
\phi(u,t)$
gives
$$\eqalignno{ a_{01}\to&{1\over \phi_u^2}a_{01},\cr
a_{02}\to&{1\over \phi_u^4}[-\phi_{uuu}\phi_u+{3\over 2}\phi_{uu}^2
                    +\phi_u^2a_{02}],\cr
a_{03}\to&a_{03},\cr
a_{04}\to&\phi_t+\phi_ua_{04}.&(6.10)\cr
}$$
Using these coordinate transformations we take $a_{04}=0$, which implies
that $a_{01}$ and
$a_{02}$
are independent of $t$, hence the evolution  equation in independent of $x$ and
$t$.
If we set $(a_{01})_u=0$, then
$$(a_{02})_{uuu}+8(a_{02})_u a_{02}=0,\eqno(6.11)$$
hence the integrable equation in this class is
$$u_t=u_3-{3\over 2}{u_2^2\over u_1}+a_{01}{1\over u_1}
+a_{02}(u)u_1^3\eqno(6.12)$$
where $a_{02}$ satisfies (6.11). It can be seen that when $a_{02}$ is
independent of $u$, it is transformed to the Callegero-Degasperis-Fokas
equation,
$$v_t=v_3-{1\over 2} v_1^3+v_1[-a_{01}e^{-2v}+3a_{02}
e^{2v}].$$
On the other hand, if $a_{02}$ is not constant it can not be
transformed to a {\it local} equation.

\vskip .2cm
\noindent
{\bf 6.3 Classification for the case $A=-{3\over
2}[u_1^2+c(x,t,u)]^{-1}u_1$. }

In this case, using the conserved density conditions we first
solve  $B$, as a function depending  on the derivatives of $c$.
Then the compatibility of the differential equations for
$C$ implies that $c_x=0$, and we can set
$c=1$ by choosing $\phi_u$. It then follows that $B=0$. As a result we obtain
$$C=a_{01}(t)(u_1^2+1)^{3/2}+a_{02}(u,t)u_1(u_1^2+1)+a_{03}(t)u_1+a_{04}(t).
$$
 We can set $a_{04}=0$ by choosing $\phi_t$, and $a_{03}=0$ by a linear
change in $x$.  Then we obtain the following equation
$$u_t=u_3-{3\over 2}
u_1(u_1^2+1)^{-1}u_2^2+a_{01}(u_1^2+1)^{3/2}+a_{02}(u)u_1(u_1^2 +1),
\eqno(6.13)$$ where
$$(a_{02})_{uuu}+8(a_{02})_ua_{02}=0,\quad\quad(a_{02})_ua_{01}=0.\eqno(6.14)$$
Thus for $a_{01}\ne 0$ we have an equation independent of $u$.

The conserved density
$\rho_0=-3(1+u_1^2)^{-1}u_1u_2$ is trivial and using the
transformation $v=\int (1+u_1^2)^{-1/2}u_2={\rm ln}\mid
u_1+\sqrt{1+u_1^2}\mid$, we obtain
$$u_t-u_3+{3\over 2}{u_1\over 1+u_1^2}u_2^2\to v_t-v_3+{1\over 3}
v_1^3.$$
Then if $a_{02}$ is constant (6.13) is transformed to
$$v_t=v_3-{1\over 2} v_1^3+{3\over4}v_1
[(a_{01}+a_{02})e^{2v}+(a_{02}-a_{01})e^{-2v}-2a_{02}].$$
However for $a_{01}=0$, $a_{02}$ need not be constant and (6.13) is not
transformable to a local equation.

\vskip .5cm

\noindent
{\bf 7. Conclusion.}
\vskip .3cm

We have shown that all evolution equations of the form (2.2) can be transformed
to an equation independent of $x$ and $t$. The surprising fact is that the
classification of equations with explicit $t$ dependence is considerably
simplified by the availibility of time dependent transformations.

It is known that integrable equations of the form $u_t=u_3+f(x,t,u,u_x)$ are
all related to the KdV equation via Miura transformations. The equations with
$u_{xx}$ dependence that are not transformable to one of these equations are
$$\eqalignno{
u_t=&u_3-{3\over 2}{u_2^2\over u_1}+a_{01}{1\over u_1}
+a_{02}(u)u_1^3\cr
  u_t=&u_3-{3\over 2}
u_1(u_1^2+1)^{-1}u_2^2+a_{02}(u)u_1(u_1^2 +1),\cr}$$
where
$$(a_{02})_{uuu}+8(a_{02})_u a_{02}=0.$$
An equivalent form of the first equation is
$$u_t=u_3-{3\over 2}{u_2^2\over u_1}+{1\over u_1}[4u^3+k_1u+k_2].$$
It has been checked that this equation do not have fifth and seventh order
local symmetries.  This result suggest the following possibilities. (i) The
formal symmetry condition will not hold at a later stage, but the formal
symmetry condition has been checked at later stages and no inconsistency has
been observed. (ii) The formal symmetry condition will hold at all orders, but
it will not be possible to write the recursion operator in closed form.
It has been shown that such a situation occurs for a system of evolution
equations [4]. Hence these two equations need further investigation.

\vskip .5cm

\noindent
{\bf Acknowledgents.}
Part of this work has been done at the University of Paderborn, under
the DFG grant. The author would like to thank  B. Fuchssteiner and W.
Oevel for valuable discussions. This work is partially supported by
the Scientific and Tecnical Research Council of Turkey.

\vskip .5cm

{\bf References.}
\baselineskip=10pt

\item{[1]} A.V. Mikhalov, A.B. Shabat and V.V Sokolov. ``The symmetry approach
to the classification
of integrable equations" in `{\it What is Integrability?} edited by
V.E. Zakharov (Springer-Verlag, Berlin 1991).
\vskip .1cm

\item{[2]} M. G\"rses and A. Karasu, ``Variable coefficient third order
Korteweg-de Vries type of equations", {\it J. Math. Phys.}, {\bf 36}, 3485,
(1995).
\vskip .1cm

\item{[3]} P.J. Olver, Lie {\it Application of Lie
Groups to Differential Equations} (Springer-Verlag, Berlin 1993)
\vskip .1cm

\item{[4]}
A.H.  Bilge, ``A system with a recursion operator but one higher
symmetry", Lie Groups and their Applications, {\bf 1} 132-139 (1994).
\vskip .1cm

\vfill
\eject

\end